\documentclass[aps,prd,preprint,showpacs,floats,nofootinbib]{revtex4-1}
\usepackage{graphicx,subfigure}
\usepackage{bm}
\usepackage{euscript,amsmath}

\begin{document}
\title{Implications of the recent measurement of pure annihilation $B_s \to \pi^+ \pi^-$ decays in QCD factorization }
\author {Guohuai Zhu}
\email[E-mail address: ]{zhugh@zju.edu.cn}
\affiliation{Zhejiang Institute of Modern Physics, Department of Physics, \\
 Zhejiang University, Hangzhou, Zhejiang 310027, P.R. China}

\date{\today}
\begin{abstract}
The CDF 3.7 sigma evidence of pure annihilation $B_s \to \pi^+ \pi^-$ decays, if confirmed, would imply a large annihilation scenario in the QCD factorization approach. This is somewhat unexpected as the large annihilation scenario was disfavored in previous studies. In this paper we reinvestigate the role of annihilation topology in QCD factorization. We find that it is not easy to reach the CDF central value of $B_s \to \pi^+ \pi^-$ decays when other decay channels are considered. Our analysis also reveals that, for well-measured charmless B decays into two final pseudoscalar mesons, the QCD factorization predictions with large annihilation parameters show good agreement with the experimental data except $B_s \to K^+ K^-$ and $B_d \to K^0 \bar{K}^0$ decays. Though other possibilities can not be excluded, this may indicate that the SU(3) flavor symmetry breaking should be taken into account for the annihilation topology. In addition, there are different annihilation topologies, so that somewhat different annihilation parameters may be chosen for different final states and different annihilation topologies. If so, the predictive power of the QCD factorization method may be rather limited for many decay channels.

\end{abstract}
\maketitle
Very recently the CDF collaboration has reported a $3.7$ $\sigma$ evidence for pure annihilation $B_s \to \pi^+ \pi^-$ decays, together with a measurement of (also pure annihilation) $B_d \to K^+ K^-$ decays \cite{CDFnote}
\begin{align}
{\cal B}(B_s \to \pi^+ \pi^-)&=(0.57 \pm 0.15 \pm 0.10) \times 10^{-6}~, \nonumber \\
{\cal B}(B_d \to K^+ K^-)&=(0.23 \pm 0.10 \pm 0.10) \times 10^{-6}~,
\end{align}
where the first errors are statistical and the second systematic. As the large hadron collider (LHC) has been running very well this year, the LHCb collaboration should be able to check and improve these results very soon.

Theoretically it was noticed first in \cite{Keum:2000ph,Keum:2000wi,Lu:2000em} that weak annihilation amplitudes may not be negligibly small in charmless B decays and was predicted in perturbative QCD method in \cite{Ali:2007ff,Li:2004ep,Li:2005vu} with the same central value as the experiment. In QCD factorization method (QCDF) \cite{Beneke:1999br,Beneke:2000ry,Beneke:2001ev,Beneke:2003zv}, although weak annihilation contributions are formally power suppressed in $\Lambda_{QCD}/m_b$, they are supposed to be important, together with the chirally-enhanced power corrections, to account for the large branching ratios of penguin-dominated B decays. In addition, the annihilation topologies may provide large strong phases which are crucial to accommodate the significant direct CP violation of $B_d \to \pi^- K^+$. Unfortunately these power correction terms are not calculable in QCDF as they contain endpoint singularities which violate the factorization theorem. Phenomenologically these chirally-enhanced power corrections and weak annihilation amplitudes are estimated in a model dependent way \cite{Beneke:2001ev}, and comprehensive studies of charmless B decays (see, e.g., \cite{Beneke:2003zv,Du:2002cf,Beneke:2006hg,Cheng:2008gxa,Cheng:2009mu,Cheng:2009cn}) show good agreement with the experiment in general. However in soft collinear effective theory (SCET) \cite{Bauer:2000yr,Bauer:2001cu,Bauer:2004tj}, it was argued that the chirally enhanced power corrections and weak annihilation diagrams are actually factorizable and real \cite{Jain:2007dy,Arnesen:2006vb}, while the so-called charming penguins \cite{Colangelo:1989gi,Ciuchini:1997hb,Ciuchini:2001gv} are supposed to be non-perturbative and important to account for the large branching ratios and CP violations of charmless B decays.

Noticed that theoretically it is still controversial on whether the charming penguins would invalidate the standard picture of QCD factorization \cite{Beneke:2009az,Beneke:2004bn}. It is also hard in practice to tell whether charming penguins are really important, as generally weak annihilation and charming penguins have the same topology \footnote{The ratio ${\cal B}(B_s \to \rho^+ K^-)/{\cal B}(B_s \to \pi^+ K^-)$ may provide some insight into this issue, as discussed in \cite{Zhu:2010eq}.}. But with the first experimental evidence of pure annihilation $B_s \to \pi^+ \pi^-$ decays, it is clear now that the annihilation contributions must be significant in charmless B decays, irrespective of whether the charming penguins are large or not.

It is therefore timely to reinvestigate the annihilation contributions in QCDF, especially considering that the annihilation amplitudes seemed to be underestimated in previous QCDF studies. For instance, ${\cal B}(B_s \to \pi^+ \pi^-)$ was estimated to be $0.155 \times 10^{-6}$ in the favored scenario $S4$ of \cite{Beneke:2003zv}, which is about three times smaller than the CDF observation. One might wonder that, with the annihilation magnitude larger than expected, whether it is still possible for the QCDF predictions of charmless B decays to be consistent with the experiments. This motivates our work below.

In the framework of QCDF, the decay amplitudes of $B_s \to \pi^+ \pi^-$ and $B_d \to K^+ K^-$ can be expressed as \cite{Beneke:2003zv}
\begin{align}
A(B_s \to \pi^+ \pi^-)&=B^s_{\pi\pi}\left (V_{ub}^\ast V_{us}\Big [ b_1+2b_4+\frac{1}{2}b_{4,EW}\Big ]+ V_{cb}^\ast V_{cs}\Big[ 2b_4+\frac{1}{2}b_{4,EW}\Big ] \right ) \nonumber \\
A(B_d \to K^+ K^-)&=B^d_{KK}\left (V_{ub}^\ast V_{ud}\Big [ b_1+2b_4+\frac{1}{2}b_{4,EW}\Big ]+ V_{cb}^\ast V_{cd}\Big[ 2b_4+\frac{1}{2}b_{4,EW}\Big ] \right )
\end{align}
with
\begin{align}
B^s_{\pi\pi}=i \frac{G_F}{\sqrt{2}} f_{B_s} f_\pi f_\pi~,\hspace*{1cm} B^d_{KK}=i \frac{G_F}{\sqrt{2}} f_{B} f_K f_K~.
\end{align}
b's are the annihilation coefficients defined as \cite{Beneke:2001ev}
\begin{align}
b_1=\frac{C_F}{N_c^2}C_1 A_1^i~, \hspace*{1cm} b_4=\frac{C_F}{N_c^2} \Big [ C_4 A_1^i + C_6 A_2^i \Big ]~,
\hspace*{1cm} b_{4,EW}=\frac{C_F}{N_c^2} \Big [ C_{10} A_1^i + C_8 A_2^i \Big ]~.
\end{align}
For the case of two pseudoscalars in the final states, $A_{1,2}^i$ are found to be
\begin{align}
A_1^i \simeq A_2^i \simeq 2\pi \alpha_s \left ( 9 \Big [ X_A-4+\frac{\pi^2}{3} \Big ] + r_\chi^2 X_A^2 \right )~,
\end{align}
where approximate SU(3) flavor symmetry has been used for $r_\chi^\pi=r_\chi^K=r_\chi \simeq 2m_K^2 /(m_b (m_q+m_s))$ and $X_A$ parameterizes
the endpoint singularity as
\begin{align}
X_A=\ln \frac{m_B}{0.5 \mbox{GeV}}\Big(1+\rho_A e^{i\phi_A} \Big)~.
\end{align}
Notice that $\phi_A$ is an arbitrary strong phase and normally $\rho_A \le 1$ is assumed, which reflects our ignorance on the
annihilation amplitudes dominated by the soft gluon interaction.

\begin{figure}
\includegraphics[scale=1.0]{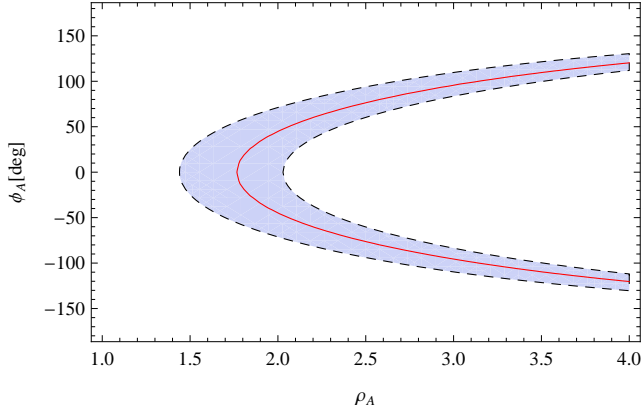} \caption{\label{fig:1} Contour plot of the branching ratio of $B_s \to \pi^+ \pi^-$ decay as a function of  the annihilation parameters $\rho_A$ and $\phi_A$. The solid red line represents the experimental central value and the light blue (grey) region
corresponds to one sigma contour. }
\end{figure}

It is then straightforward to estimate the pure annihilation B decays in QCDF. Taking
\begin{align}\label{eq:input1}
f_{B_s}=230~\mbox{MeV}~, \hspace*{1cm} m_b(m_b)=4.2~\mbox{GeV}~, \hspace*{1cm} m_s(2~\mbox{GeV})=80~\mbox{MeV}~,
\end{align}
and the Wolfenstein parameters \cite{Charles:2004jd}
\begin{align}\label{eq:CKM}
A=0.812~, \hspace*{1cm} \lambda=0.2254~, \hspace*{1cm} \bar{\rho}=0.144~, \hspace*{1cm} \bar{\eta}=0.342~,
\end{align}
we show in Fig. \ref{fig:1} the contour plot of ${\cal B}(B_s \to \pi^+ \pi^-)$ as a function of the parameters $\rho_A$ and $\phi_A$.
One may observe from Fig. \ref{fig:1} that $\rho_A$ must be larger than $1.5$ to account for the measured branching ratio of $B_s \to \pi^+ \pi^-$ within one sigma experimental error. Notice that we had chosen a relatively small s quark mass here which can enhance the penguin amplitudes, as
adopted in the favored scenario S4 of \cite{Beneke:2003zv}. The large $\rho_A$ scenario has been discussed in \cite{Beneke:2003zv} and concluded to be unlikely as a fine-tuning of strong phase $\phi_A$ is required to satisfy the experimental bounds on $\pi K$, $\pi K^\ast$ and $\rho K$ systems. But the experimental evidence of $B_s \to \pi^+ \pi^-$ forced us to have a closer look on the large annihilation scenario in the following.

\begin{figure}
\subfigure[]{
\includegraphics[width=0.45\textwidth]{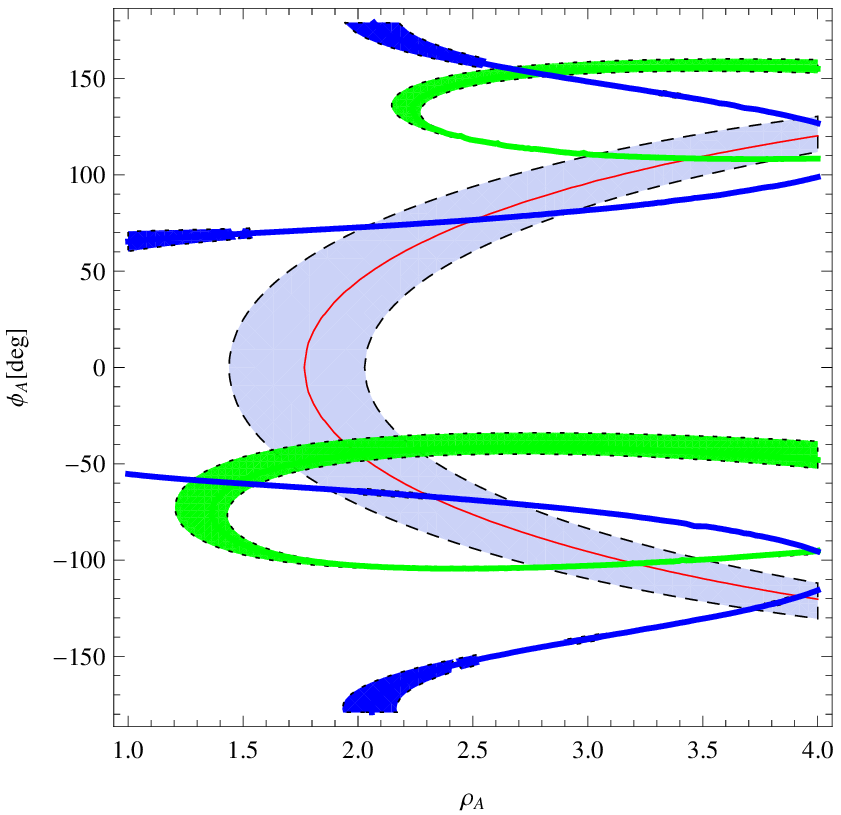}} \hspace*{0.2cm}
\subfigure[]{
\includegraphics[width=0.45\textwidth]{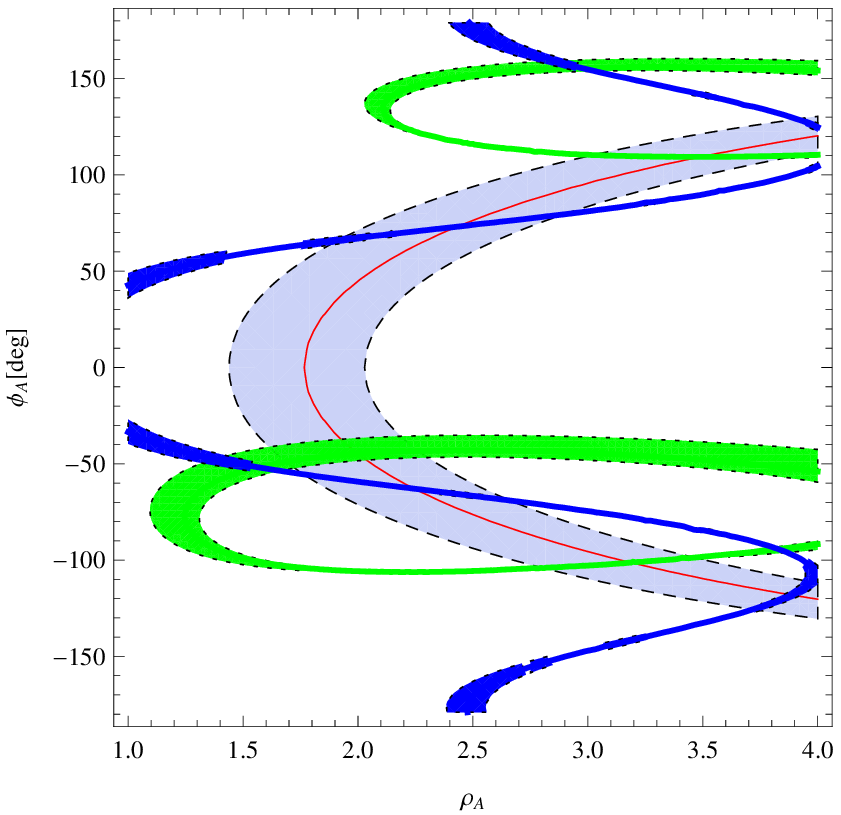}}
\caption{\label{fig:2} Contour plots of the branching ratio and direct CP violation of $B_d \to \pi^- K^+$ channel, as well as the branching ratio of $B_s \to \pi^+ \pi^-$, as functions of the annihilation parameters $\rho_A$ and $\phi_A$. The blue and green bands represent the branching ratio and direct CP violation of $B_d \to \pi^- K^+$ decay, respectively, within two sigma experimental errors. The meaning of solid red line and the light blue region is the same as in Fig. \ref{fig:1}. The form factor $F^{B\pi}$ is taken to be $0.26$ (left plot) or $0.22$ (right plot). }
\end{figure}

In the QCDF method, nonzero $\phi_A$ in the annihilation amplitudes could provide large strong phase which is also required to explain the measured direct CP violation of charmless B decays. Concerning the $B_d \to \pi^- K^+$ channel which has been well measured \cite{Barberio:2008fa}
\begin{align}
A_{CP}(B_d \to \pi^- K^+)=(-9.8^{+1.2}_{-1.1})\%~, \hspace*{1cm} {\cal B}(B_d \to \pi^- K^+)=(19.4 \pm 0.6) \times 10^{-6}~,
\end{align}
one may determine, together with the $B_s \to \pi^+ \pi^-$ constraint, the annihilation parameters $\rho_A$ and $\phi_A$ as shown in Fig. \ref{fig:2}. The form factor $F^{B\pi}$ brings an important source of uncertainty to $B_d \to \pi^- K^+$ decay, which has been estimated in light-cone sum rules as $0.26 \pm 0.03$ \cite{Ball:2004ye,Duplancic:2008ix}. But as shown in the left plot of Fig. \ref{fig:2}, there is no overlap between different bands if the central value of $F^{B\pi}$ is taken. This means there is no solution of $\rho_A$ and $\phi_A$ which can satisfy simultaneously the experimental constraints of $B_d \to \pi^- K^+$ and $B_s \to \pi^+ \pi^-$ decays. Instead, a small (but still reasonable) form factor $F^{B\pi}=0.22$ helps to reconcile the QCDF predictions with observations, as can be seen in the right plot of Fig. \ref{fig:2}. One may also observe from Fig. \ref{fig:2} that, considering $B_d \to \pi^- K^+$ constraints, it is hard in QCDF to obtain ${\cal B}(B_s \to \pi^+ \pi^-)$ as large as the CDF central value $0.57 \times 10^{-6}$, though it is possible to reach the lower side of the $1~\sigma$ error band.
In the following we will take the parameters
\begin{align}\label{eq:anni-para}
 \rho_A^{PP}=1.75~,\hspace*{0.7cm} \phi_A^{PP}=-53^\circ~, \hspace*{0.7cm} F^{B\pi}=0.22~,\hspace*{0.7cm} F^{BK}=0.28~,\hspace*{0.7cm} F^{B_sK}=0.26~,
\end{align}
where slightly small form factors $F^{BK}$ and $F^{B_sK}$ have also been adopted, compared with the light-cone sum rules estimation $0.33 \pm 0.04$ \cite{Ball:2004ye} and $0.30^{+0.04}_{-0.03}$ \cite{Duplancic:2008tk}, respectively. The superscript "PP" in the above equations means these annihilation parameters are adopted only for hadronic B decays into two light pseudoscalar mesons in the final states. For the parameters of wave functions which have less impacts on our results, we simply take \cite{Beneke:2003zv,Ball:2006wn}
\begin{align}\label{eq:other-input}
\lambda_B=200~\mbox{MeV}~,\hspace*{0.7cm} a_2^\pi=0.25~,\hspace*{0.7cm} a_1^K=0.06~,\hspace*{0.7cm} a_2^K=0.25~.
\end{align}
We shall use the input parameters listed in Eqs.(\ref{eq:input1},\ref{eq:CKM},\ref{eq:anni-para},\ref{eq:other-input}) as an illustration to check the QCDF predictions in comparison with data for some selected decay channels.

\begin{table}
\begin{tabular}{|l|c|l|l|c|l|}
\hline
~~~~~~~Mode & QCDF & Experiment & ~~~~~~~~Mode & QCDF & Experiment \\
\hline
${\cal B}(B_s \to \pi^+ \pi^-)$ & $0.40$ & $0.57 \pm 0.18$ &
${\cal B}(B_d \to K^+ K^-)$ & $0.20$ & $0.23 \pm 0.14$ \\
\hline
${\cal B}(B_d \to \pi^- K^+)$ & $20.6$ & $19.4 \pm 0.6$ &
$A_{CP}(B_d \to \pi^- K^+)$ & $-11.4$ & $-9.8^{+1.2}_{-1.1}$ \\
\hline
${\cal B}(B^+ \to \pi^0 K^+)$ & $12.5$ & $12.9 \pm 0.6$ &
$A_{CP}(B^+ \to \pi^0 K^+)$ & $-2.3$ & $5.0 \pm 2.5$ \\
\hline
${\cal B}(B_d \to \pi^0 K^0)$ & $9.4$ & $9.5 \pm 0.5$ &
${\cal B}(B^+ \to \pi^+ K^0)$ & $23.2$ & $23.1 \pm 1.0$ \\
\hline
${\cal B}(B_s \to \pi^+ K^-)$ & $7.1$ & $5.0 \pm 1.1$ &
$A_{CP}(B_s \to \pi^+ K^-)$ & $37.4$ & $39 \pm 17$ \\
\hline
${\cal B}(B_s \to K^+ K^-)$ & $45.1$ & $26.5 \pm 4.4$ &
${\cal B}(B^+ \to \pi^+ \pi^0)$ & $5.8$ & $5.9 \pm 0.4$ \\
\hline
${\cal B}(B_d \to \pi^+ \pi^-)$ & $6.1$ & $5.2 \pm 0.2$ &
$A_{CP}(B_d \to \pi^+ \pi^-)$ & $29.7$ & $38 \pm 6$ \\
\hline
${\cal B}(B^+ \to K^+ \bar{K}^0)$ & $1.6$ & $1.36^{+0.29}_{-0.27}$ &
${\cal B}(B_d \to K^0 \bar{K}^0)$ & $1.9$ & $0.96^{+0.21}_{-0.19}$ \\
\hline
\end{tabular}
\caption{\label{table:1} CP-averaged branching ratios (in unit of $10^{-6}$) and direct CP asymmetries (in units of $10^{-2}$) of some $B \to PP$ decay modes in the framework of QCDF, with the input parameters displayed in Eqs.(\ref{eq:input1},\ref{eq:CKM},\ref{eq:anni-para},\ref{eq:other-input}).}
\end{table}

Our results are listed in Table \ref{table:1} for the CP-averaged branching ratios and direct CP asymmetries of some $B \to PP$ decay channels. Notice that we have not discussed hadronic B decays with final states containing $\eta$ or $\eta^\prime$, as there are additional large uncertainties in QCDF relating to the relevant flavor-singlet components. For the decay mode $B^+ \to \pi^0 K^+$, naively one might expect $A_{CP}(\pi^- K^+) \sim A_{CP}(\pi^0 K^+)$ which is in disagreement with the experiments by $5.3 \sigma$. So we have followed \cite{Cheng:2009eg,Cheng:2009cn} to adopt the scenario of large color-suppressed tree topology which may arise from spectator scattering or final state interactions. This $K\pi$ puzzle may also be explained by the so-called Pauli blocking effect proposed recently by Lipkin \cite{Lipkin:2011hh}.

\begin{figure}
\includegraphics[scale=1.0]{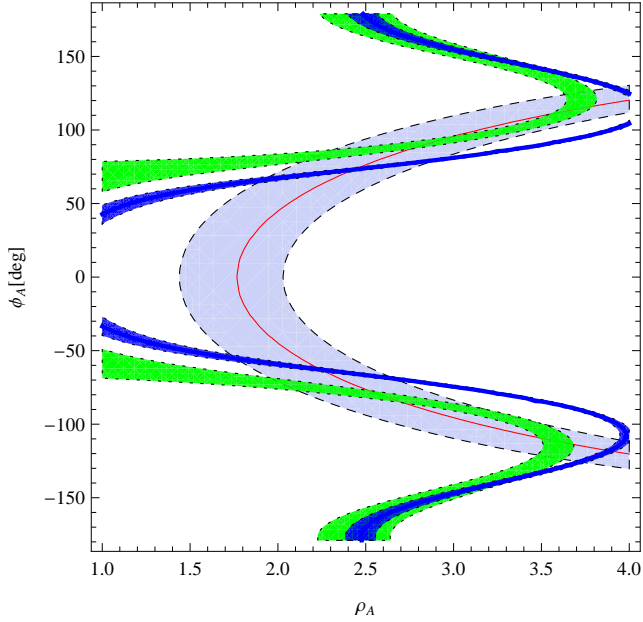} \caption{\label{fig:3} Contour plots of the branching ratios of $B_d \to \pi^- K^+$, $B_s \to K^+ K^-$ and $B_s \to \pi^+ \pi^-$ decays, as functions of the annihilation parameters $\rho_A$ and $\phi_A$. The blue and green bands represent the branching ratios of $B_d \to \pi^- K^+$ and $B_s \to K^+ K^-$ decays, respectively, within one sigma error. The meaning of solid red line and the light blue region is the same as in Fig. \ref{fig:1}.  }
\end{figure}

The QCDF results in Table \ref{table:1} show good agreement with the experiments in general, except $B_s \to K^+ K^-$ and $B_d \to K^0 \bar{K}^0$ decays, whose branching ratios are estimated to be nearly twice larger than the experimental data. Notice that these two decay channels have something in common: both of them are penguin dominated with the annihilation amplitudes determined essentially by a combination of $b_3+2b_4$. For the other penguin-dominated modes listed in Table \ref{table:1}, namely $B \to \pi K$ channels, the annihilation amplitudes are largely determined by $b_3$ term while pure annihilation decay $B_s \to \pi^+ \pi^-$ is dominated by $b_4$ term. We show in Fig. \ref{fig:3} the contour plots of ${\cal B}(B_d \to \pi^- K^+)$, ${\cal B}(B_s \to K^+ K^-)$ and ${\cal B}(B_s \to \pi^+ \pi^-)$ in the plane of annihilation parameters $\rho_A-\phi_A$. One may observe that there is no overlap between different bands, which means there is no solution in the $\rho_A-\phi_A$ plane, with other parameters fixed, to reproduce these three decay channels in agreement with the data simultaneously. For instance, one may take
\begin{align}\label{eq:alter-anni}
\rho_A^{PP}=2~,\hspace*{1cm} \phi_A^{PP}=-70^\circ~,
\end{align}
to get
\begin{align}
{\cal B}(B_s \to K^+ K^-)=29.6 \times 10^{-6}~,\hspace*{1cm}{\cal B}(B_s \to \pi^+ \pi^-)=0.4 \times 10^{-6}~,
\end{align}
which are within one sigma error of the experimental data. But with the same parameters we obtain
\begin{align}
{\cal B}(B_d \to \pi^- K^+)=12.6 \times 10^{-6}~,\hspace*{1cm}A_{CP}(B_d \to \pi^- K^+)=-20.7\%~,
\end{align}
which are in strong contradiction with the experimental results $(19.4\pm0.6) \times 10^{-6}$ and $-9.8^{+1.2}_{-1.1}\% $, respectively.

We have tried to vary some sensitive parameters to find a solution. For instance, one may raise the form factor $F^{B_s K}$ to have less tension with constraints of Fig. \ref{fig:3}, but then ${\cal B}(B_s \to \pi^+ K^-)$ will become too large. One may instead take a larger form factor $F^{B\pi}$ but as we have shown in Fig. \ref{fig:2}(a), it can not work when $A_{CP}(B_d \to \pi^- K^+)$ is included. We have also tried to vary the s quark mass from $80$ MeV to $95$ MeV but it does not help to reduce the discrepancy either.

In QCDF calculations, we have followed the common practice to assume universal annihilation parameters $\rho_A$ and $\Phi_A$ for all $B_{d,s} \to PP$ decay modes, which is respected in the limit of SU(3) flavor symmetry. However, the SU(3) breaking effects could be as large as of $O(20\%)$. Observing that the difference of the annihilation parameters between Eq.(\ref{eq:anni-para}) and Eq.(\ref{eq:alter-anni}) is just about $20$ percent, one possible way out of this problem is to introduce SU(3) breaking effects into annihilation parameters. That is to say, somewhat different $\rho_A$ and $\phi_A$ may be introduced for hadronic B decays into different final states.

In the annihilation diagrams, the gluons may emit either from initial-state parton (denoted as $A^i_{1,2,3}$) or from final-state parton (denoted as $A^f_3$). As a common practice, the annihilation parameters $X_A$ has been assumed to be universal for both annihilation topologies. But it is possible that $X_A$ in $A^f_3$ is different from that in $A^i_{1,2,3}$, as they originate from different topologies. Observing that $B_s \to \pi^+ \pi^-$ depends only on $A^i_{1,2,3}$, while $B \to \pi K$, $K^0 \bar{K}^0$ and $B_s \to K^+ K^-$ decays contain both annihilation topologies but with different expressions, the above mentioned disagreement may also be solved by assuming $X_A^f$ to be somewhat different from $X_A^i$. As the SU(3) breaking effects of ${\cal O}(20-30\%)$ should be considered in general, it could be that $X_A$'s are different for different decay channels and different annihilation topologies.

In any case, as the QCDF predictions are very sensitive to the annihilation parameters, the predictive power in the framework of QCDF may be rather limited for many charmless decay channels. Notice also that, as there is no SU(3) flavor relation between $B \to PP$ and $B \to PV,~VV$ decays, the large annihilation scenario in $B \to PP$ decays does not necessarily mean that $\rho_A$ should also be around $2$ in $B \to PV$ and $VV$ decays.

In summary, the first CDF evidence of $B_s \to \pi^+ \pi^-$ decays, which could be verified very soon by the LHCb collaboration, implies a large annihilation scenario with $\rho_A$ around $2$ in the QCDF method. This is surprising as previous studies in QCDF preferred $\rho_A \simeq 1$.
So we checked in details whether the large annihilation scenario is consistent with the experimental data for many well-measured $B \to PP$ decay modes. Considering $B_d \to \pi^- K^+$ constraints, we observed that it is hard in QCDF to obtain ${\cal B}(B_s \to \pi^+ \pi^-)$ as large as the CDF central value $0.57 \times 10^{-6}$, though it is possible to reach the lower side of the $1~\sigma$ error band. We found in addition that, taking slightly smaller form factors than the light cone sum rules estimation, the QCDF predictions are in good agreement with the data in general, except $B_s \to K^+ K^-$ and $B_d \to K^0 \bar{K}^0$ decays whose branching ratios are predicted to be almost twice larger than the experimental measurements. One possible way to solve this problem is to take into account the SU(3) breaking effects in the annihilation parameters, which however means that the predictive power is rather limited in the QCDF method.

\section*{Acknowledgement}

This work is supported in part by the National Science Foundation of China (No. 11075139 and No.10705024).  G.Z is also supported in part by
the Fundamental Research Funds for the Central Universities.

\end{document}